\documentclass[aps,twocolumn,prb,groupedaddress,showpacs]{revtex4}

\usepackage[utf8]{inputenc}
\usepackage{graphicx,color}

\def\im{{{i}}}
\newcommand{\op}[1]{{\hat #1}}
\newcommand{\dt}{\Delta\tau}

\begin{document}


\title{Orbital-selective Mott Transitions in a Doped Two-band Hubbard Model}


\author{Eberhard Jakobi}
\email[Email: ]{eberhard.jakobi@uni-mainz.de}
\author{Nils Bl\"umer}
\author{Peter van Dongen}
\affiliation{Institute of Physics, Johannes Gutenberg University, 55099 Mainz, Germany}


\date{\today}

\begin{abstract}
We extend previous studies on orbital-selective Mott transitions in the paramagnetic state of the half-filled degenerate two-band Hubbard model to the general doped case, using a high-precision quantum Monte Carlo dynamical mean-field theory solver. For sufficiently strong interactions, orbital-selective Mott transitions as a function of total band filling are clearly visible in the band-specific fillings, quasiparticle weights, double occupancies, and spectra. The results are contrasted with those of single-band models for similar correlation strengths.
\end{abstract}

\pacs{{71.30.+h},{71.27.+a},{71.10.Hf},{71.10.Fd}}

\maketitle


\section{Introduction}
Starting with the early theoretical work by Hubbard\cite{hubbard_hIII64} and Brinkman and Rice,\cite{PhysRevB.2.4302} and experimental work on transition metal oxides,\cite{PhysRevB.7.326,56312321654} the Mott-Hubbard metal-insulator transition (MIT) has been a subject of great interest in condensed matter physics for decades.
Much of its essential physics is captured already by the single-band Hubbard model,\cite{gutzwiller:HubbardModel63,kanamori:HubbardModel,hubbard_hIII64} as has been shown in numerous studies using the dynamical mean-field theory (DMFT) framework.\cite{PhysRevLett.62.324,georges:dmft96}
Multi-band extensions of the Hubbard model allow for a more realistic description of the MIT and of other strong-correlation phenomena and have, in general, much richer phase diagrams.
Multi-orbital Mott transitions have been explored (within DMFT) in a doped two-band model with {\it equivalent} orbitals already more than 10 years ago.\cite{PhysRevB.55.R4855}
In this case, the band degeneracy increases the number of Mott lobes from 1 to 3, but has no fundamental impact on the Mott physics near half filling (in contrast, e.g., to ferromagnetism  for which the inter-orbital Hund exchange is essential\cite{PhysRev.49.537,RevModPhys.25.220,PhysRevB.56.3159,PhysRevB.57.6896,b8243}).

More recently, it has been realized that orbital degeneracy can change the character of Mott transitions in a fundamental way: namely, in the case of multiple {\it inequivalent} orbitals. In the electronic context, such inequivalence can arise naturally by orbital-dependent hopping amplitudes, associated, e.g., with in-plane versus out-of-plane $t_{2g}$ orbitals in layered ruthenates.\cite{PhysRevLett.84.2666,PhysRevB.62.6458} It has been suggested by Anisimov et al.\cite{EPJB.25.191} that such systems should undergo a sequence of orbital-selective Mott transitions (OSMTs) with increasing interaction strength, which would then explain the peculiar phase diagram of Ca$_{2-x}$Sr$_x$RuO$_4$.\cite{PhysRevLett.84.2666,PhysRevB.62.6458} Clearly, such a scenario -- with coexisting itinerant and localized valence electrons in the intermediate orbital-selective Mott phase -- is fundamentally different from a conventional simultaneous Mott transition of all valence electrons.

Within the last few years, it has been established that OSMTs occur in multi-band Hubbard models with inequivalent orbitals under quite general circumstances.\cite{JPhysCondMat.19.436206,inaba:155106,de'medici:205124,PhysRevB.72.205126,PhysRevLett_99_236404,liebsch:116402,inaba:085112,koga:216402,PhysRevLett.91.226401,osmt_0506151}
A (nearly) minimal Hamiltonian for orbital-selective Mott behavior is the two-band Hubbard model with band-specific hopping amplitudes $t_m$ (for orbital index $m\in\{1,2\}$) and Ising type Hund rule couplings (parametrized by $J_z$),
\begin{eqnarray}
\op H &=& -\sum_{\langle ij\rangle m\sigma} t_{m}^{\phantom{\dagger}} \op c_{im\sigma}^\dagger \op c_{jm\sigma}^{\phantom{\dagger}}+ U\sum_{im}\op n_{im\uparrow}^{\phantom{\dagger}}\op n_{im\downarrow}^{\phantom{\dagger}}+\nonumber \\
\label{eq:model} &\,&\sum_{i\sigma\sigma'}\left(U' -\delta_{\sigma\sigma'}^{\phantom{\dagger}}J_z^{\phantom{\dagger}}\right)\op n_{i1\sigma}^{\phantom{\dagger}}\op n_{i2\sigma'}^{\phantom{\dagger}}\, .
\end{eqnarray}
Here, $\sigma\in\{\uparrow,\downarrow\}$ denotes the spin; $i$ and $j$ label lattice sites; $\op n_{im\sigma}\equiv \op c_{im\sigma}^\dagger \op c_{im\sigma}^{\phantom{\dagger}}$. In contrast to the hopping, the intra-orbital Hubbard interaction $U$ is assumed orbital independent. The intra- and interorbital interactions are related by $U = U' + 2J_z$. In the following, we will refer to Eq.\ (\ref{eq:model}) as the $J_z$-model.\cite{knecht2005,PhysicaBCondMat.359.1366,pvd2007}

A general discussion of the Hund exchange would have to include spin-flip and pair hopping terms,
\[
\op H_\perp^{\phantom{\dagger}} = \frac{1}{2} J_{\perp}^{\phantom{\dagger}}\sum_{im\sigma} \op c_{im\sigma}^\dagger \left( \op c_{i\bar{m}\bar{\sigma}}^\dagger \op c_{im\bar{\sigma}}^{\phantom{\dagger}} +\op c_{im\bar{\sigma}}^\dagger \op c_{i\bar{m}\bar{\sigma}}^{\phantom{\dagger}} \right) \op c_{i\bar{m}\sigma}^{\phantom{\dagger}}\, .
\]
However, such terms are not essential for OSMTs (Ref.\ \onlinecite{knecht2005})
and change their character only in the $SU(2)$ symmetric limit
$J_z=J_{\perp}\equiv J$ (i.e., in the Heisenberg limit). In
particular, it was shown for the half-filled case\cite{PhysRevLett.95.206401} 
that the itinerant electron
species in the orbital-selective Mott phase is a non-Fermi-liquid
for all $J_\perp<J_z$ and a Fermi liquid only at $J_\perp = J_z$. In this sense,
$J_z$ model (\ref{eq:model}) represents the generic class of anisotropic Hund couplings.

The nature of OSMTs and their essential requirements are well understood for particle-hole symmetric situations. This symmetry is destroyed upon doping (i.e., by variation of the chemical potential $\mu$) and/or by adding an orbital dependent field, i.e., by crystal field splitting.
Koga et al.\ studied effects of weak hole doping on OSMTs in  a two-band Hubbard model.\cite{koga:216402} Crystal field effects have been investigated in two-orbital\cite{werner:126405} and three-orbital\cite{0611075} systems. In a ferromagnetic slave-boson mean-field theory calculation, R\"uegg et al.\ \cite{23987239857} obtained a phase diagram for the two-band Hubbard model with crystal field splitting, which includes ferromagnetic as well as band insulating regimes.

In this paper, we extend high-precision\cite{PhysRevB_76_205120} quantum Monte Carlo (QMC) calculations\cite{knecht2005,pvd2007,353542354} within DMFT for two-band model (\ref{eq:model}) to general doping levels. Of central interest is the phase diagram, which is derived from orbital-dependent fillings, double occupancies, and quasiparticle weights. At the same time, these observables as well as spectra are used for characterizing the orbital-selective physics, i.e., the effects of orbital inequivalence.  Other results of interest in this paper include
unexpected behavior in the intermediate interaction regime, where the double occupancy for the wide band becomes nearly independent of the interaction. We also show
that the entire doping range $0\leq n\leq 4$ is subdivided into three
different regimes by the transport behavior of the narrow band,  and
that the two-band spectra differ from analogous single-band results in that
the additional scattering channels introduced by the interorbital
couplings tend to smear out the spectrum. Finally, we will discuss the relation between orbital-selective Mott physics and non-Fermi-liquid behavior, i.e., whether these phenomena always occur simultaneously (in the $J_z$ model).
Here, as in previous work on this subject, we concentrate on the effect of correlations on the thermodynamic and spectral properties of the system in the paramagnetic phase, i.e., excluding magnetism.

The structure of this paper is as follows: In Section \ref{subsec:methods} we briefly discuss the methods and model parameter values relevant to our work. Section \ref{subsec:observables} gives information on the observables studied in our paper. Static and dynamic observables are treated separately (in Secs.\ \ref{sec:static} and \ref{sec:dynamic}, respectively). The static observables discussed in Sec.\ \ref{sec:static} include band-resolved particle numbers (Sec.\ \ref{subsec:filling}), intraorbital double-occupancies (Sec.\ \ref{subsec:double}), and the phase diagram as a function of interaction and total filling (Sec.\ \ref{subsec:phase}). As dynamic observables, we study band-resolved spectral functions (Sec.\ \ref{subsec:spectra}), the spectral weight at the Fermi edge (Sec.\ \ref{subsec:N0}), and the Matsubara self-energy as well as quasiparticle weights (Sec.\ \ref{subsec:QP}). Finally, in Sec.\ \ref{summary}, we summarize the results and formulate our conclusion.


\subsection{\label{subsec:methods}Methods}
In the following, we consider $J_z$-model (\ref{eq:model}) for two bands with a bandwidth ratio of 2. For the narrow band, we assume a semi-elliptic ``Bethe'' non-interacting DOS with a fixed hopping amplitude $t_\mathrm{n}=0.5$ and, hence, a full bandwidth $W_\mathrm{n} = 2$. Similarly, we assume a semi-elliptic DOS with $t_\mathrm{w}=1$ and $W_\mathrm{w} = 4$ for the wide band. The intraband on-site interaction $U$ is chosen as the primary (variable) interaction parameter; it acts equally in both orbitals. The interorbital interactions are scaled as $J_z = U/4$ and $U' = U/2$, in line with most earlier studies.\cite{23987239857,knecht2005,inaba:155106,PhysRevLett_99_236404} This work is restricted to the paramagnetic case; spin indices are omitted when appropriate. We choose a fixed temperature $T=1/40$, which is of the order of the critical temperature in the undoped case.\cite{pvd2007}

Within DMFT, two-band model (\ref{eq:model}) is mapped to a two-orbital single-impurity Anderson model which has to be solved self-consistently.\cite{georges:dmft96} In this work, the impurity model is solved using the Hirsch-Fye quantum Monte Carlo (HF-QMC) algorithm.\cite{PhysRevLett.56.2521,PhysRevLett.69.168} This method discretizes the imaginary-time path integral expression for the Green function into $\Lambda$ time slices of uniform width $\dt=\beta/\Lambda$, where $\beta=1/T$ (for $k_{\text{B}}\equiv 1$); a Hubbard-Stratonovich (HS) transformation replaces the electron-electron interaction at each time step by a binary auxiliary field which is sampled using standard Markov Monte Carlo techniques. The results are exact in the combined limit of vanishing discretization $\dt\to 0$ and a large number of Monte Carlo update sweeps. In this work, the systematic errors associated with finite discretization ($\dt=0.4$, unless noted otherwise) are minimized by supplementing the discrete QMC Green function with a high-frequency expansion of the self-energy.\cite{knechtMasterthesis,nilsPhd} The high precision of this method has proved essential in detecting both OSMTs in the undoped case;\cite{knecht2005} if required, extrapolations $\dt\to 0$ can be used in order to achieve extreme precision\cite{PhysRevB.71.195102} and efficiency.\cite{PhysRevB_76_205120} In order to compute spectral functions, the imaginary-time Green functions resulting from QMC are analytically continued to the real axis using a standard maximum-entropy method (MEM).\cite{234523452345345}

We have verified that the remaining discretization bias (at $\dt=0.4$) is generally small (much smaller than in preceding QMC implementations\cite{PhysRevB.51.10411}), up to slight shifts in the critical interactions, by performing additional simulations with $0.25\le \dt \le 0.5$. In particular, we did not find significant effects on the single-particle spectra: their general shapes as well as peak positions and weights were unchanged.


\subsection{\label{subsec:observables}Observables}
Using the QMC-DMFT methods, described above, we investigate several observables in this paper: band-specific particle numbers and intraorbital double occupancies, the density of states, in particular its value at the Fermi level, and the quasiparticle weight. We now briefly introduce these quantities, which are of interest by themselves and are used in order to construct the phase diagram.

To start with the general properties of $J_z$-Hamiltonian (\ref{eq:model}): this model is formulated in a particle-hole symmetric way (in particular, no crystal field splitting is included), so that results for a total filling of $n > 2$ can be easily mapped to results for $n < 2$. As a result of this well-known symmetry, particle- and hole-dopings are equivalent and, accordingly, the word ``doping'' in the following refers to both. Moreover, the phase diagrams, to be calculated below, will be mirror symmetric with respect to the half-filled case, where $n=2$.

The study of orbital-specific particle numbers and intraorbital double occupancies is particularly interesting for parameter values, for which the orbitals behave physically differently, i.e., in the orbital-selective Mott phase (OSMP). In this case, the narrow band is in the insulating state and displays a gap, while the wide band is in a non-Fermi-liquid metallic state. The insulating nature of the narrow band physically suggests that both the densities and the double occupancies of this band may initially remain pinned upon doping, and in fact one of the goals of our investigation of these observables is to determine, whether and to what extent (depending upon the interaction strength $U$) this is true.

Clearly, in addition to the transport properties (e.g., the electrical conductivity), which are not explicitly investigated in this paper, one of the most fundamental observables for detecting metal-insulator transitions is the density of states (DOS), i.e., the local spectral function. It is particularly interesting to contrast the results for the DOS in the two-band calculations with single-band results for comparable parameter sets, whenever appropriate.

Another important indicator for the onset of insulating behavior is the quasiparticle weight $Z$, which is mathematically determined by the self-energy $\Sigma(\omega)$. Upon approaching the insulating phase of a {\it single-band\/} model from the metallic side, the quasiparticle weight vanishes (at least in the ground state) at the metal-insulator transition. On account of the relation $Z = m/m^*$, the vanishing of the quasiparticle weight is equivalent to the divergence of the effective mass $m^*$. It is important to note that the physical interpretation of $Z$ and $m^*$ as the ``quasiparticle weight'' and the ``effective mass'', respectively, is meaningful only in a Fermi-liquid regime. Within a Fermi-liquid phase, the quasiparticle weight is well approximated in an imaginary-time calculation by its imaginary-time analog:
\begin{eqnarray}
Z = \frac{m}{m^*} &\equiv & \left( 1-\frac{d \mathrm{Re}\Sigma}{d\omega}\bigg|_{\omega = 0} \right)^{-1} \nonumber \\
\label{eq:Z}&\simeq& \left[ 1+\frac{\mathrm{Im}\Sigma(\im\omega_1)}{\pi T} \right]^{-1}\, .
\end{eqnarray}
Here $\omega_1 = \pi T$ is the first Matsubara frequency. Both expressions for the quasiparticle weight clearly recover the non-interacting limit $Z\rightarrow 1$ at arbitrary temperatures for $\Sigma\rightarrow 0$. In addition, the imaginary-time secant approximation is exact in a Fermi-liquid phase in the limit $T\rightarrow 0$. However, both expressions are demonstrably distinct, e.g., in the insulating phase at $T=0$, since the real-frequency definition of $Z$ vanishes, whereas estimate (\ref{eq:Z}) remains finite. This illustrates that estimate (\ref{eq:Z}) will in general be inaccurate or even invalid in non-Fermi-liquid regimes, such as in particular the OSMP, where the narrow band is insulating, while the wide band is a non-Fermi-liquid type metal (see). In any case, the discrete imaginary-time expression in Eq.\ (\ref{eq:Z}) is an established measure of the low-frequency behavior of the self-energy.


\section{\label{sec:static}Static observables as a function of doping}
The total electron concentration in a two-band model is composed of contributions from the two orbitals: $n = n_\mathrm{w}+n_\mathrm{n}$, where $n_\mathrm{w}=
\sum_\sigma\langle \op n_{i\mathrm{w}\sigma}\rangle$ and $n_\mathrm{n}=
\sum_\sigma\langle \op n_{i\mathrm{n}\sigma}\rangle$ represent the particle filling of the wide and the narrow band, respectively.

In two-band system (\ref{eq:model}) with total half filling, due to particle-hole symmetry, either band is exactly half-filled: $n_\mathrm{w} = n_\mathrm{n} = 0.5$.  Let us briefly review the central results obtained earlier for this special case:\cite{knecht2005} with increasing interaction, and starting from the non-interacting limit (and for low $T=1/40$), both bands undergo  consecutive Mott transitions. For weak interaction ($U<U_\mathrm{c1} \approx 2.1$), both bands remain itinerant. For strong interaction ($U>U_\mathrm{c2}\approx 2.6$), both bands become insulating, since the energy difference between the atomic levels becomes comparable to the interaction $U$. In the intermediate region, the wide band is metallic and the narrow band is insulating. Slight variations in temperature do not significantly change the results quantitatively.\cite{pvd2007}
This picture agrees with arguments in Ref.\ \onlinecite{EPJB.25.191}, relating the interacting two-band model to two distinct bands, which separately undergo the Mott transition according to their individual bandwidth.

In the following, we first discuss our results for the band-specific particle numbers and double occupancies, which are then compared to predictions from a simple rigid-band model and from perturbation theory. In comparing the QMC results to simple model calculations, the doping dependence turns out to be of great interest. We also present a phase diagram for the doped two-band model, with particular emphasis on transport characteristics.
Accordingly, we distinguish insulating, orbital-selective Mott and metallic phases.

\vspace*{-3mm}\subsection{\label{subsec:filling}Band-specific particle numbers}
\begin{figure}
\includegraphics[angle=270,scale=0.5]{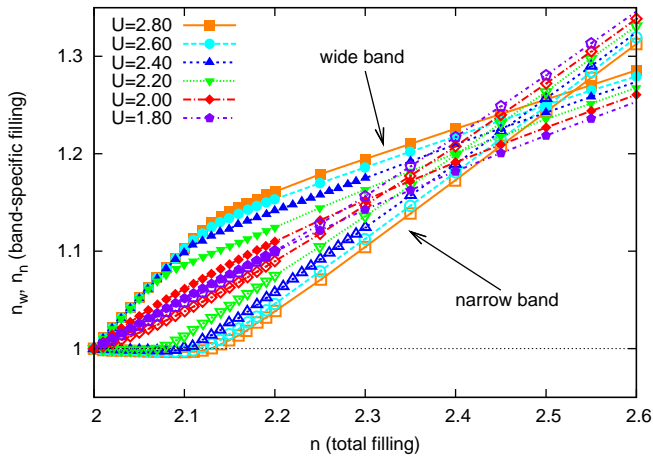}\\
\caption{(Color online) Band-specific fillings for the two-band model versus
total filling for a range of interactions $U$. Open and solid
symbols represent the narrow and the wide band, respectively.
}\label{fig:filling}
\end{figure}

We start the presentation of our results with the band-specific fillings, which are plotted in Fig.\ \ref{fig:filling} for different interaction strengths $1.8\le U\le 2.8$. 
For small deviations from half filling and $U>2$, only the wide band accounts for the doping, while the narrow band remains (to a very good approximation) half-filled. As a consequence, the relation  $n_\mathrm{w} \simeq n - 1$ holds approximately in this low-doping regime. As one expects, the pinning of the particle filling of the narrow band at its half-filled value in the low-doping regime becomes more extended for increasing $U$. These plateaus are bounded by well-localized kinks, beyond which the curves $n_\mathrm{n}(n)$ show rapid, nearly linear, increase: additional electrons are predominantly allocated to the narrow band. Hence, the slope of $n_\mathrm{w}(n)$ drops significantly. The corresponding phase boundary will be discussed below, in Sec.\ \ref{subsec:phase}.
Note that the curves in Fig.\ \ref{fig:filling} for the narrow and the wide bands {\it cross\/} in the OSMP regime near $n\simeq 2.4$, with details depending upon $U$.


\subsection{\label{subsec:double}Double occupancies}
\begin{figure}
\includegraphics[angle=270,scale=0.5]{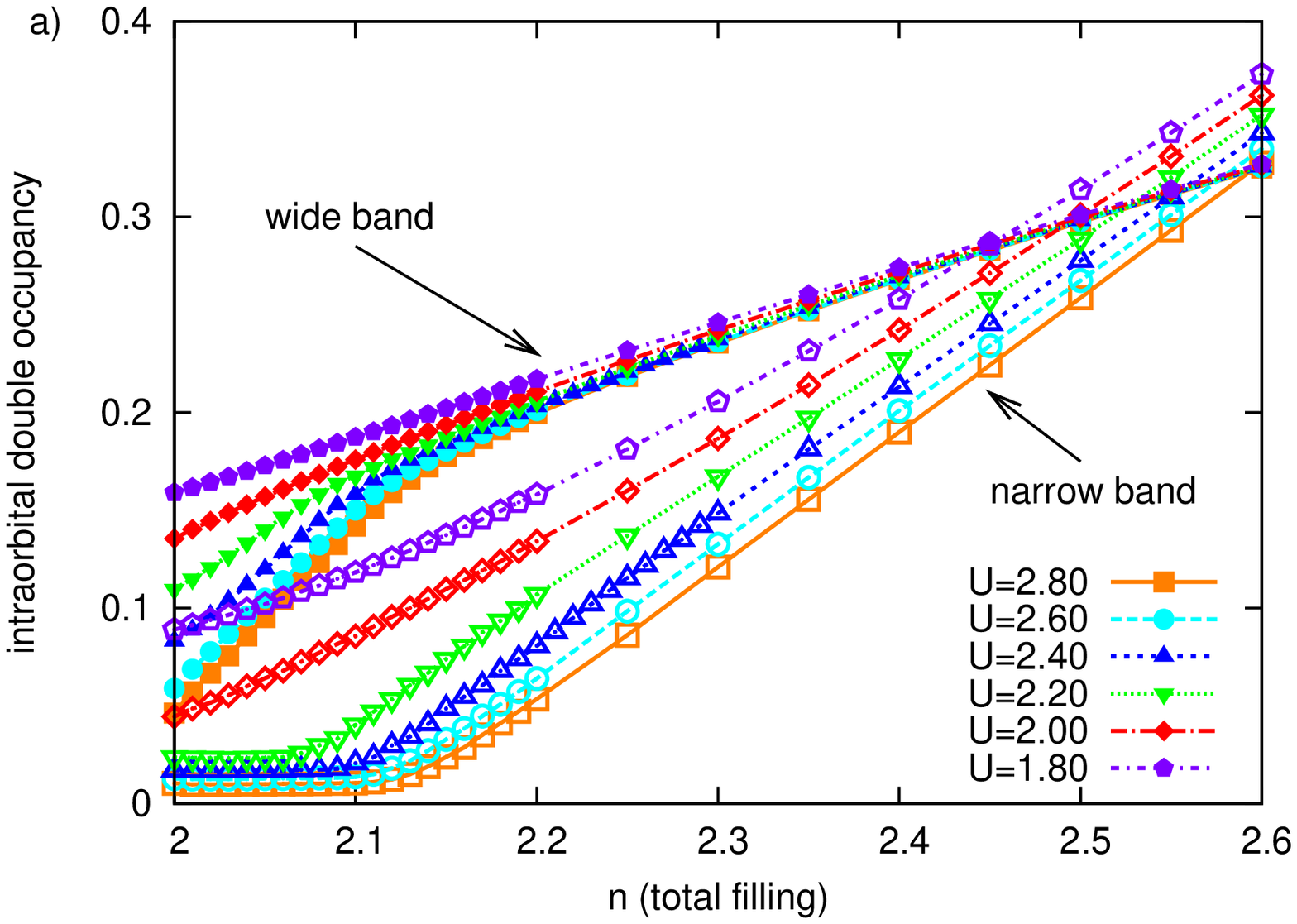}\\
\includegraphics[angle=270,scale=0.5]{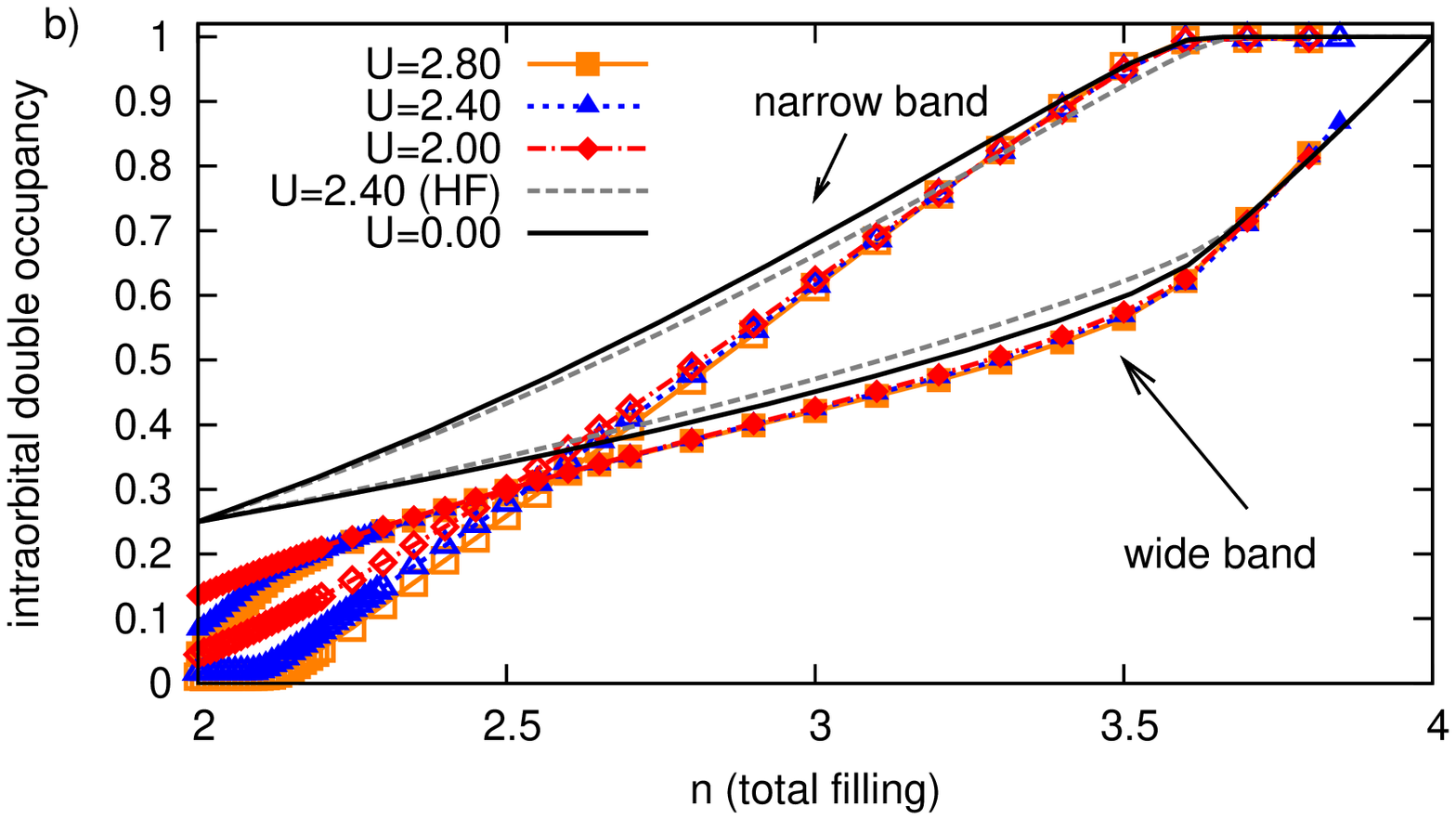}
\caption{(Color online) Intraorbital double occupancy for various $U$-values as a function of doping for the two-band model.
Panel ($a$) shows for dopings near half filling and panel ($b$) in the entire doping range $2\leq n\leq 4$ that the intraorbital double occupancy for the wide band becomes nearly independent of the interaction for intermediate $U$ and all fillings $n\gtrsim 2.2$. In ($b$), results for the noninteracting case ($U=0$, black line) and within Hartree-Fock approximation (for $U=2.4$, long-dashed line) are included for comparison.
}\label{fig:double}
\end{figure}

In Fig.\ \ref{fig:double}a, we show QMC results for the intraorbital double occupancies $\langle \op n_{im\uparrow} \op n_{im\downarrow}\rangle$ ($m\in\{w,n\}$) of the wide band (upper curves) and the narrow band (lower curves). In the weakly doped regime ($n<2.1$) of the OSMP (with $U\gtrsim 2.2$), the intraorbital double occupancy for the narrow band - which, as was demonstrated above, is effectively singly occupied - is strongly suppressed. In this regime ($n\lesssim 2.1$), the double occupancy of the wide band rapidly decreases with increasing interaction due to Coulomb repulsion. In the stronger doped regime ($n\gtrsim 2.1$)
the situation reverses: The double occupancy is supressed in the narrow band with increasing interaction, whereas the intraorbital double occupancy for the wide band becomes nearly $U$ independent. As we will see below, both bands are metallic in this regime ($n\gtrsim 2.1$) and show a significant DOS in the vicinity of the Fermi edge, the DOS of the narrow band obviously being larger than that of the wide band. As a consequence, both the average filling and the concentration of doubly occupied orbitals increases more rapidly in the narrow band than in the wide band for  $n\gtrsim 2.1$. Hence, as for the particle numbers, a crossing of the curves for the wide and narrow bands occurs at $n\simeq 2.5$, so that the concentration of intraorbital double occupancies is {\it larger\/} for the narrow than for the wide band for $n\gtrsim 2.5$. Moreover, these curves show a very weak $U$-dependence in this regime, which suggests that very similar behavior is to be expected from weak-coupling perturbation theory. We will come back to this issue below. The crossing and the virtual $U$ independence can be clearly seen in Fig.\ \ref{fig:double}b, which shows the QMC results for the double occupancies in the entire regime $2\leq n\leq 4$; for comparison, the corresponding weak-coupling (Hartree-Fock) and noninteracting results are also shown.

\begin{figure}[t]
\includegraphics[angle=270,scale=0.5]{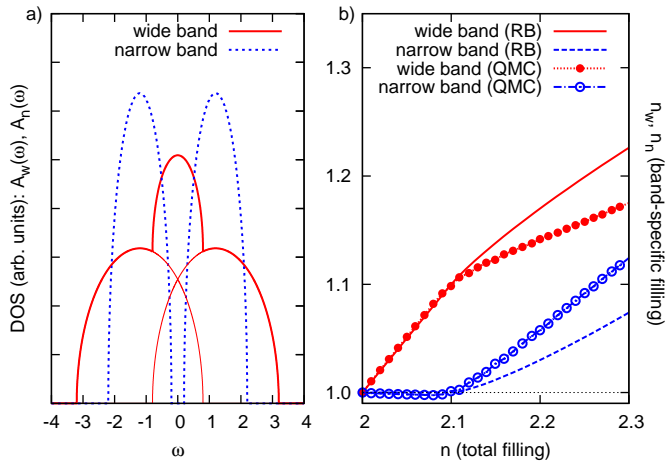}
\caption{(Color online) ($a$) DOS of a rigid-band model with a semi-elliptic DOS and bandwidths $W_\mathrm{n} = 2$ and $W_\mathrm{w} = 4$, representative of interactions $U \simeq 2.4$. ($b$) Band-specific fillings within this model, compared to QMC results for $U = 2.4$ and $T=1/40$.
}\label{fig:rigid}
\end{figure}

The primary mechanism governing the band-specific fillings and the intraorbital double occupancies in the {\it low-doping\/} regime can be understood on the basis of a rigid-band model with two bands and intraorbital interaction only. A comparison between the band-specific fillings within this rigid-band model [sketched in Fig.\ \ref{fig:rigid}] 
with the QMC results reveals good agreement for $n\lesssim 2.1$, so that the pinning of the narrow-band filling can be explained as resulting from the single-particle gap. If one further assumes that the narrow-band double occupancy is a function of the narrow-band filling and of the interactions, it has also to remain pinned in the same regime -- in full agreement with the QMC results.
 The validity of the rigid-band model rapidly breaks down as the total filling is increased beyond $n>2.1$, where the actual particle numbers of the narrow (wide) band increase more rapidly (slowly) than in the rigid-band model. The deviations between the simple rigid-band assumption and the true results in the metallic phase ($n>2.1$) are due to correlations, in particular to the formation of Kondo resonances in both bands.

\begin{figure}
\includegraphics[angle=270,scale=0.5]{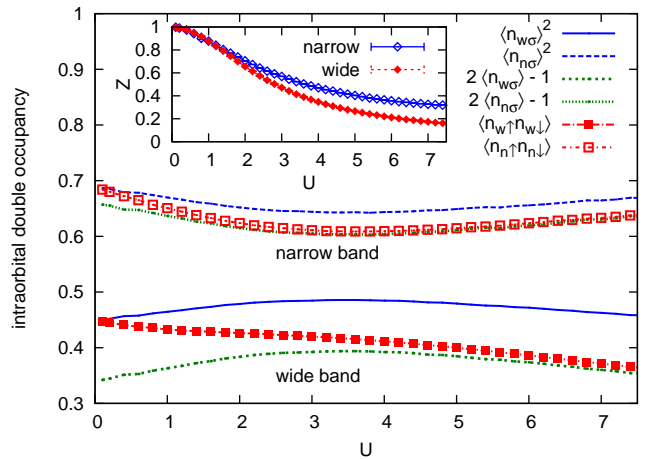}
\caption{(Color online) Intraorbital double occupancy versus $U$ for density $n = 3$ (squares). In the interaction range $2.5 \lesssim U\lesssim 5$, the intraorbital double occupancy is only very weakly dependent on $U$. Density-dependent upper and lower bounds on the double occupancies are obtained in the limits of weak (dashed/solid lines) and strong (dotted lines) correlation, respectively. Inset: quasiparticle weight $Z$ (cf. Sec.\ \ref{subsec:QP}).
}\label{fig:doubleZ}
\end{figure}

\begin{figure}[t]
\includegraphics[angle=270,scale=0.5]{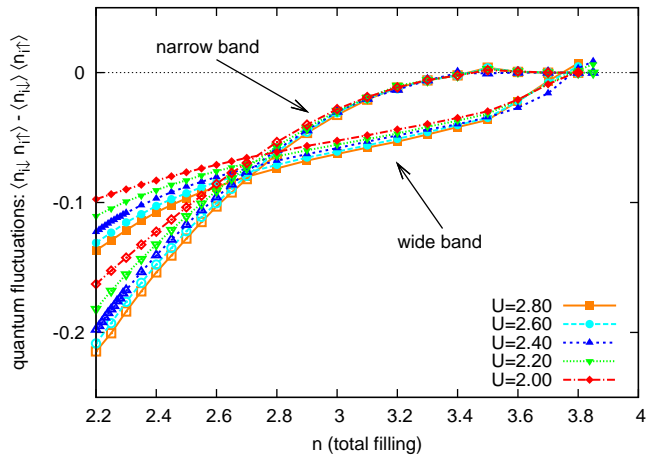}
\caption{(Color online) The $\uparrow\downarrow$-density-density correlation function for total fillings $n\gtrsim 2.2$, corresponding to quantum fluctuations in the intraorbital double occupancy around the uncorrelated limit.
The correlations are generally {\it negative\/}, since an enhanced $\uparrow$-density on a site implies a decreased $\downarrow$-density. At fixed interaction $U$, the amplitude of the correlation function {\it decreases\/} monotonically with filling, since the available phase space for density fluctuations decreases with increasing $n>2$.
}\label{fig:corr}
\end{figure}

We now discuss and interpret the remarkably weak $U$-dependence of the intraorbital double occupancies of the narrow and the wide band for $n\gtrsim 2.5$.
Before comparing to perturbative results, we illustrate the weak dependence of the double occupancies on $U$ in Fig.\ \ref{fig:doubleZ}, which covers the interaction regime $0\leq U\lesssim 7$ for a total density $n=3$. Figure \ref{fig:doubleZ} indeed shows a near-constancy of the double occupancies for such interaction values. On the basis of the virtual $U$-independence, in particular for $2\lesssim U\lesssim 4$, one expects good agreement between QMC-results and perturbation theory in this regime. As can be seen from Fig.\ \ref{fig:double}b, where corresponding Hartree-Fock results are also plotted,
this is indeed what one finds. Figure \ref{fig:double}b shows that the wide band agrees with perturbation theory better than the narrow band, at least for $n\lesssim 3$, as one expects, since the effective interaction strength (compared to the band width) is larger for the narrow than for the wide band. This is also confirmed by Fig.\ \ref{fig:corr}, which shows the quantum fluctuations around the Hartree-Fock solution, which (for $n\lesssim 2.7$) are indeed largest for the narrow band. The near-perfect agreement between the narrow-band QMC results and perturbation theory for $n\gtrsim 3.2$ follows immediately from the near absence of singly occupied or empty orbitals in the narrow band in that regime. For completeness we should add that the weak $U$-dependence of the concentrations of intraorbital double occupancies clearly cannot persist for arbitrarily large $U$ values. From strong-coupling perturbation theory one expects that, for sufficiently large $U$, the double occupancies of {\it both\/} bands converge towards a common strong-coupling value $0.5$.

In between such a strong-coupling phase and the metallic weak-coupling phase, one would expect a metal-insulator transition. As mentioned in the introduction, such additional Mott lobes (at quarter and three quarter filling) had been found in an early HF-QMC study of a two-band Hubbard model, even with critical interactions similar to those at half filling.\cite{PhysRevB.55.R4855} However, this study considered the case of SU(4) symmetric interactions ($U=U'$ and $J=0$) and equal bandwidth. The first difference is particularly relevant, since the Heisenberg exchange strongly suppresses Mott phases at (three) quarter filling in terms of the critical interactions: the phase boundaries are shifted to much larger $U_c$. This is easily seen by comparing the cost of a charge fluctuation in the atomic limit ($\frac{1}{4} U$ at $n=1$ or $n=3$, compared to $\frac{7}{4} U$ at $n=2$; both for $J=U/4, U'=U/2$) and has also been found numerically for SU(2) symmetric interactions.\cite{0706-1.3948} In the present case, we expect the system to remain metallic (at $n=3$) up to $U_c\gtrsim 20$, i.e., beyond the interaction range easily accessible using QMC calculations. In addition, the critical temperatures may be considerably lower; however, sharp crossover lines should persist at least up to the temperatures studied in this paper.\cite{Koga05PRB,Gorelik09}

From the data of Fig.\ \ref{fig:doubleZ}, we can, indeed, exclude a phase transition at $n=3$ for $U\lesssim 7$: both the intraorbital double occupancies (squares, main panel) and the quasiparticle weights $Z$ (inset) are completely smooth as a function of the interaction $U$, i.e., do not show any kinks or hysteresis effects that would be characteristic of an MIT. Note that both the double occupancy and $Z$ are larger for the narrow band due to its larger filling fraction.

\subsection{\label{subsec:phase}Phase diagram}
Next we discuss the phase diagram as a function of the interaction strength $U$ and the total filling $n$ with particular emphasis on transport-related properties.
To do so, we distinguish an {\it insulating\/} state of the two-band system from an {\it orbital-selective Mott\/} phase, in which one band is metallic and the other insulating, and a purely {\it metallic\/} phase, in which neither of the bands is insulating. As is well-known, all three phases occur already at half filling, where the system is in the metallic, orbital-selective Mott or insulating state for $U\leq U_{\rm{c1}}\simeq 2.1$, $U_{\rm{c1}}\leq U\leq U_{\rm{c2}}\simeq 2.6$, or $U_{\rm{c2}}\leq U$, respectively.

For the doped system ($n\neq 2$) it is already clear from our investigation of particle numbers and of double occupancies for the {\it narrow\/} band at low doping, which appeared to be pinned at the half-filled state for moderate-to-large $U$-values, that this band is {\it insulating\/} in this regime. Since the particle filling of the wide band changes substantially with total density (according to $n_\mathrm{w}\simeq n-1$) for these $(U,n)$-values, one expects the wide band to be metallic in this part of the phase diagram. Accordingly, one can take the pinning of the narrow-band particle numbers as a criterion for the occurrence of an OSMP. Alternatively, at larger doping, where the fillings of both bands change upon changing the total density, it is clear that both bands have finite compressibility and, hence, are {\it metallic\/}. Finally, at non-integer total filling in the paramagnetic phase, one does not expect the occurrence of a rigorously insulating state for both bands simultaneously, since at least one of the bands must have finite compressibility in that region. Thus, to summarize, we found a simple criterion for the determination of the paramagnetic phase diagram as a function of interaction and doping: If the system shows pinning for the narrow band away from half filling, it is in an OSMP, if not, then it is metallic.

It is clear that this criterion for the determination of the phase diagram, which is essentially based on the behavior of the compressibility as a function of $U$ and $n$, should be cross-checked with alternative criteria, based on dynamical quantities, such as, e.g., the finiteness of the DOS at the Fermi level for the wide and the narrow bands. Results for such dynamical quantities, which are presented below, turn out to be fully consistent with the present ``static'' criteria and lead to the same phase diagram.

\begin{figure}
\includegraphics[angle=270,scale=0.5]{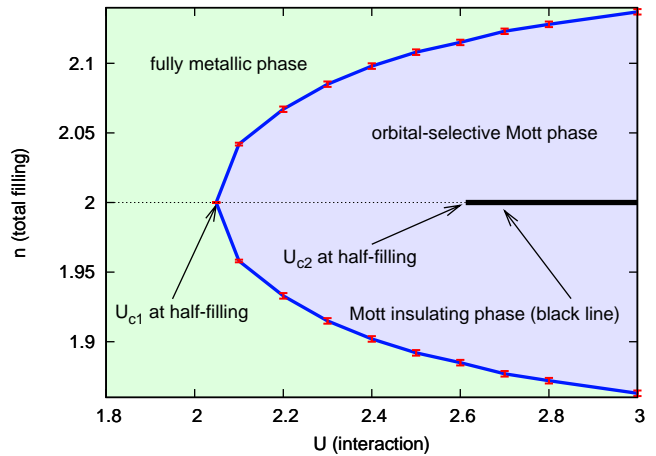}
\caption{(Color online) Orbital-selective Mott phase diagram of the two-band model for particle densities $1.85\leq n\leq 2.15$. For interaction $U < U_{\mathrm{c1}}$, both bands are metallic at all densities. The insulating phase at half filling (thick black line) only exists for $U>U_{\mathrm{c2}}$.
}\label{fig:phase}
\end{figure}

The phase diagram, which is based on the above-mentioned criteria concerning the band-specific filling factors $n_\mathrm{n}(n)$ and $n_\mathrm{w}(n)$ as a function of the total density, is presented in Fig.\ \ref{fig:phase} for the moderately strong interactions $U\le 3$, near the OSMP previously established at half filling. One finds that the purely insulating state, which occurs only at half filling as explained above, is embedded in an OSMP, which reaches from $U\simeq 2.1$ to $U=\infty$ and is itself embedded in a region of purely metallic states. As can be seen from Fig.\ \ref{fig:phase}, these metallic states occur, for a given fixed interaction strength, only for sufficiently large doping. The QMC phase diagram, presented here, is in qualitative agreement with the results of R\"uegg et al.\ ,\cite{23987239857} which were obtained based on variational methods. As a remark we add that, since, in the OSMP at {\it half filling\/}, the wide band is in a non-Fermi-liquid state,\cite{353542354} by continuity one also expects a non-Fermi-liquid state in the OSMP at low doping. Signatures of such a non-Fermi-liquid state away from half filling will indeed be found and discussed below.

Our phase diagram Fig.\ \ref{fig:phase} is in reasonable general agreement with the ground state phase diagram obtained by Inaba and Koga\cite{0706-1.3948} in the SU(2) symmetric limit (also for $J=U/4$): the critical interactions at half filling are only slightly larger in the latter case (by about $10\%$); however, at strong interactions $U\gtrsim 3$, the OSM phases seem to extend over larger filling ranges for SU(2) symmetric Hund rule couplings. Unfortunately, the coarse filling grid employed in Ref.\ \onlinecite{0706-1.3948} does not allow for a detailed comparison at the scale of Fig.\ \ref{fig:phase}. Still, we may conclude that the phase boundaries do not sensitively depend on the specifics of the Hund rule couplings (with or without spin-flip and pair hopping terms).

\begin{figure}
\includegraphics[angle=270,scale=0.5]{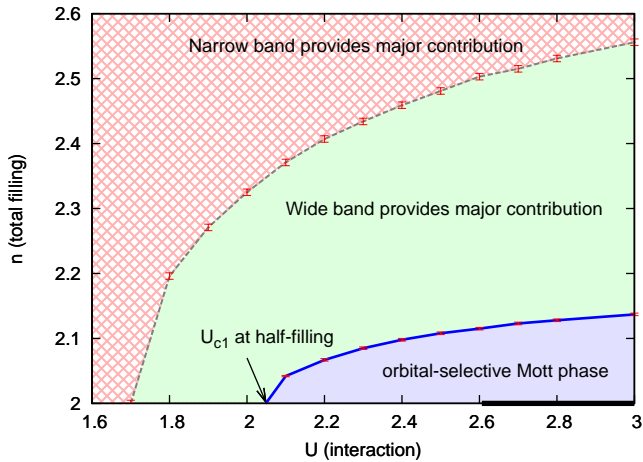}
\caption{(Color online) Phase diagram of the two-band model for densities $2\leq n\leq 2.6$. Within the range of parameter values, for which both bands are metallic, the solid green (or patterned red) area indicates that the wide (or narrow) band contributes stronger to the doping, implying $n_\mathrm{w}>n_\mathrm{n}$ (or $n_\mathrm{n}>n_\mathrm{w}$). At the boundary (dashed line), both bands are equally filled.
}\label{fig:phase2}
\end{figure}

The purely metallic phase in Fig.\ \ref{fig:phase} can be subdivided into two regimes, depending on which of the two orbitals is stronger doped (cf.\ Fig.\ \ref{fig:filling}), as shown in Fig.\ \ref{fig:phase2}. Evidently, the OSMP is entirely embedded in a region in which the wide band hosts the larger fraction of the doping, which is itself embedded in a parameter region in which the narrow band contains the larger doping fraction. The boundary between both sections of the metallic phase is marked by a dashed line, which corresponds to the crossing points of the density curves for the narrow and the wide band at fixed interaction of Fig.\ \ref{fig:filling}.
Interestingly, it is quite easy to predict the large-$U$ asymptotics of the (dashed) cross-over line: it should converge to $n=3$ for $U\to \infty$. The reason is that equal filling of both orbitals is asymptotically expected in all Mott phases, in particular also at three quarter filling (with $U_c\gtrsim 20$, cf.\ Sec.\ \ref{subsec:double}). Thus, the cross-over line should either continue, beyond $U_c$, as the Mott phase transition line or converge towards that line for $U>U_c$.


\section{\label{sec:dynamic}Dynamic observables as a function of doping}
In this section, we discuss dynamic observables: first the spectral function (at general frequencies), then, in particular, the spectral weight at the Fermi level and the functional dependence of the $Z$-factor, which corresponds to the quasiparticle weight in a Fermi-liquid phase but loses this interpretation in the OSMP or an insulating state. The spectral function and the spectral weight at the Fermi level are calculated using the maximum entropy method (MEM). Finally, the extent of non-Fermi-liquid behavior will be analyzed on the basis of imaginary-time self-energies.


\subsection{\label{subsec:spectra}Spectral function}
Both in the case of the wide band and for the narrow band a comparison to typical single-band behavior is helpful for understanding the doping dependence of the spectral function.

\begin{figure}
\includegraphics[angle=270,scale=0.5]{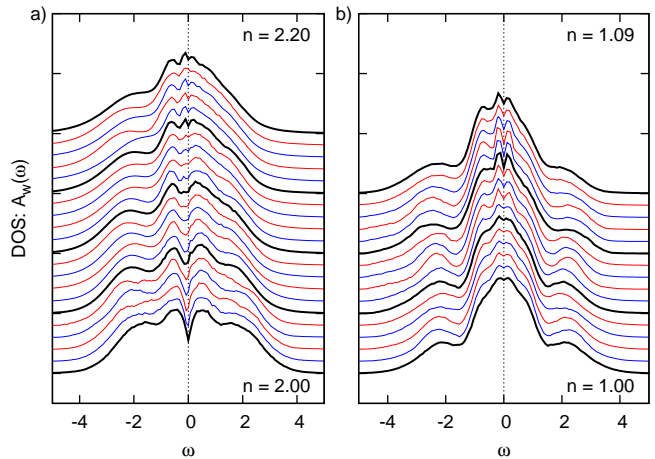}
\caption{(Color online) DOS calculated for various particle densities near half filling as a function of frequency ($a$) for the wide band in the two-band model with interaction $U=2.4$, and ($b$) for a single-band calculation with identical parameters (bandwidth $W = 4$, interaction $U=2.4$, and temperature $T = 1/40$).
}\label{fig:spectraW}
\end{figure}

The spectral function of the wide band of the two-band Hubbard model, as calculated with the MEM from our quantum Monte Carlo data for the single-particle Green function, is presented in Fig.\ \ref{fig:spectraW}a (left panel), where it is juxtaposed with the spectral function of a single-band model with the same parameters [see Fig.\ \ref{fig:spectraW}b]; the latter corresponds to the wide band in two-band model (\ref{eq:model}) where the interorbital couplings $U', J_z$ have been removed. The spectral function of the wide band in Fig.\ \ref{fig:spectraW}a is characteristic of the typical spectrum of the wide band within the OSMP. Minor changes in the interaction strength $U$ cause only slight changes in the spectrum. Comparing the result of the wide band of a two-band system to the spectral function from a single-band calculation, one finds that the spectral weight of the wide band has a broader distribution. This can be attributed to an increased scattering amplitude in the two-band model, due to the larger number of interaction channels. Moreover, the Kondo resonance is more pronounced in the single-band case and the dips between the resonance and Hubbard bands are clearly visible in the low-doping regime. One also notices a dip in the spectral function of the wide band at the Fermi edge in the low-doping domain. This phenomenon is well-known from previous work\cite{knecht2005} and can be attributed to non-Fermi-liquid behavior. In fact, the same kind of dip is found in the Hubbard-III solution or, equivalently, in the exact solution of the half-filled Falicov-Kimball model, which describes a non-Fermi liquid. Note that the dip persists to about $n=2.1$, i.e., the boundary of the OSM phase (at $U=2.4$); beyond that density, it cannot be distinguished from the (inevitable) numerical noise in the spectra. We will further discuss the extent of non-FL features in Sec.\ \ref{subsec:QP}.

We now consider the spectral function of the {\it narrow\/} band in a two-band system, which is again juxtaposed with the spectrum of a single-band model at comparable parameter values. For this purpose, we define scaled interaction strengths $\tilde U^\mathrm{(1)}\equiv U/U_\mathrm{c}^\mathrm{(1)}$ and $\tilde U^\mathrm{(2)}\equiv U/U_\mathrm{c}^\mathrm{(2)}$ for the single- and the two-band system, respectively. Here $U_\mathrm{c}^\mathrm{(1)}$ and $U_\mathrm{c}^\mathrm{(2)}$ are the critical interactions for the Mott transition of the single-band model and the metal-OSMP transition of the two-band system, respectively. The numerical values of $U_\mathrm{c}^\mathrm{(1)}$ and $U_\mathrm{c}^\mathrm{(2)}$ are approximately given by:
\begin{eqnarray*}
U_\mathrm{c}^\mathrm{(1)} \approx 2.35\hspace{5mm}&\ &\mbox{(single-band\ model)} \\
U_\mathrm{c}^\mathrm{(2)} \approx 2.10\hspace{5mm}&\ &\mbox{(two-band\ model)}
\end{eqnarray*}
The parameter values of the narrow band and the single-band system are referred to as ``comparable'' if $\tilde U^\mathrm{(1)}\approx \tilde U^\mathrm{(2)}$; then, the systems have about the same ``distance'' from a Mott transition at half filling. In actual QMC calculations it has to be taken into account that the critical interactions $U_\mathrm{c}^\mathrm{(1)}$ and $U_\mathrm{c}^\mathrm{(2)}$ depend slightly on the Trotter discretization. Moreover, in comparing the two bands, a slight mismatch of the available $\tilde U$ values is unavoidable, since the required $U_\mathrm{c}$ values are only approximately known and the ratios $\tilde U$, resulting from the simulations, are known only on in general non-identical grids.

Note that the need of rescaling arises for the interactions since the relevant parameter is $\tilde U -1$, which is strongly affected even by small changes in $U_\mathrm{c}$ in the phase region of interest. In contrast, due to the small difference of only about $10\%$ between $U_\mathrm{c}^\mathrm{(1)}$  and $U_\mathrm{c}^\mathrm{(2)}$, a rescaling of frequencies is not necessary for the comparison (but would make quantitative comparisons with other data more difficult). Similarly, it is not expected that temperature variations of the order of $10\%$ would visibly alter the spectra; consequently the comparison is made at identical (unscaled) temperatures. It should also be noted that a rescaling would not have been appropriate for Fig.\ \ref{fig:spectraW}, where the wide-band of the two-band system is compared with a single-band system of equal bandwidth $W=4$; in that case, only the two-band system is close to an MIT (of the narrow band) while the single-band system remains metallic at least for $U\lesssim 4.7$.

\begin{figure}
\begin{center}
\hspace*{4mm}\includegraphics[angle=0,scale=0.45,trim=3mm 91mm 3mm 10mm,clip]{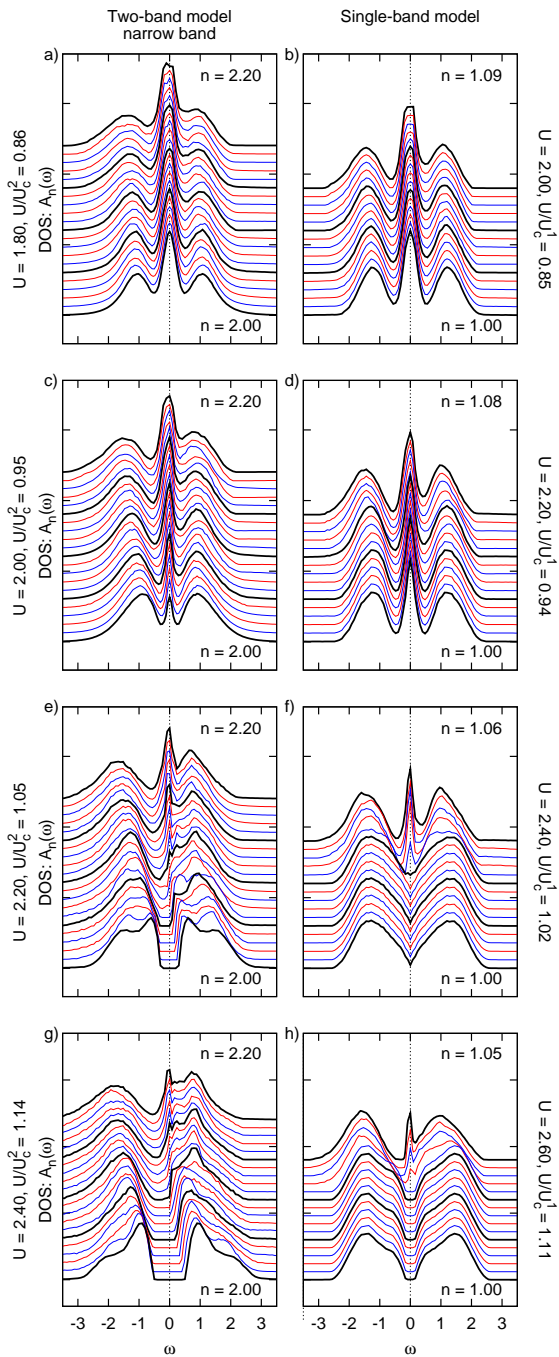}
\end{center}
\caption{(Color online) DOS calculated for various particle densities near half filling as a function of frequency, both for the narrow band in the two-band model (left column) and for a single-band calculation (right column). Panels within one row have approximately matching scaled interactions $\tilde U$. In the two-band case (left column), the {\it total} filling is varied in the range $2\leq n\leq 2.2$ in uniform steps of 0.01, with a corresponding offset.
In the single-band case (right column), the filling is varied nonuniformly: here, the offset is chosen so that adjacent spectra with the same offset have (roughly) the same {\it partial} filling (i.e., $n_{\text{n}}$ in the two-band model equals $n$ of the single-band model).
}\label{fig:spectraN}
\end{figure}

Due to the occurrence of a metal-insulator transition, the changes in the spectrum of the {\it narrow\/} band, as the interaction is varied through $\tilde U =1$, are far more drastic than in the case of the wide band. For the narrow band, therefore, we discuss results for several interaction ratios $\tilde U$, both below and above the metal-insulator transition, which are presented in Fig.\ \ref{fig:spectraN} (panels $a$, $c$, $e$, and $g$), where they are juxtaposed with the ``comparable'' single-band spectra of panels $b$, $d$, $f$, and $h$.

For interaction values $\tilde U\simeq 0.85$, clearly below the metal-insulator transition(s), we conclude from Fig.\ \ref{fig:spectraN} (panels $a$ and $b$) that the spectra of the narrow band and the single-band model show only minor differences. The Kondo resonance is well-established in both cases, and the overall shape of the spectra is very similar.
The dips in the spectrum between the resonance and the Hubbard bands are clearly more pronounced in the single-band case, in particular for larger doping.

For a relative interaction strength $\tilde U\simeq 0.95$, still slightly below the metal-insulator transition, we note from Fig.\ \ref{fig:spectraN} (panel $c$ and $d$) that, close to half filling, the Kondo resonance of the narrow band in a two-band model has considerably smaller weight and width than that of the single-band model, much beyond possible scaling effects. For the narrow band, more spectral weight is shifted from the vicinity of the Fermi level to the shoulders of the Hubbard bands so that the dips between the resonance and the Hubbard bands are, again, less pronounced for the narrow band than for the single-band system and the peaks of the Hubbard bands are closer to the Fermi level in panel $c$ than in panel $d$.

For $\tilde U\simeq 1.02-1.05$, both the narrow band of the two-band system and the single band are in a Mott insulating state at half filling, as can be seen from the gapped spectra at the lowest doping levels in panel $e$ and $f$ of Fig.\ \ref{fig:spectraN}. As the doping concentration is increased, the spectral weight near the Fermi level is larger for the single than for the narrow band, in accordance with our previous observations concerning pinning of the particle numbers of the narrow band at not-too-large doping. In spite of the pinning of the particle density ($n_{\mathrm{n}} \simeq 1$), we note from Fig.\ \ref{fig:spectraN}$e$ that the shape of the Hubbard bands of the narrow-band spectrum experiences some considerable changes at low doping. At larger doping, we note that the gap between resonance and Hubbard bands is much deeper in the single-band case than for the narrow band. Moreover, the height of the Kondo resonance of the narrow band at the largest available doping levels seems to be reduced compared to the single-band case due to a larger number of available scattering channels. The evolution of the overall shapes with increasing doping can also be characterized in the following way: in the single-band case, the quasiparticle peak grows essentially within the gap; in the two-band case, the narrow-band resonance evolves from a shoulder of the (upper) Hubbard band.

For interaction values $\tilde U\simeq 1.11-1.14$, significantly above the metal-insulator transition, one sees from panels $g$ and $h$ of Fig.\ \ref{fig:spectraN} that the behavior, already indicated in panels $e$ und $f$, becomes more pronounced. For larger doping ($n\ge 2.10$) the upper Hubbard band crosses the Fermi level, and a Kondo resonance evolves in both the narrow and the single band. For the narrow band, the resonance is strongly surpressed, however, and the vicinity of the Fermi level is dominated by the upper Hubbard band. Also, there is not a clear gap between the resonance and the lower Hubbard band of the narrow band spectrum. In contrast, the single-band model develops a very sharp Kondo resonance which is clearly separated from the lower Hubbard band by a (pseudo) gap.


\subsection{\label{subsec:N0}Spectral weight at the Fermi level}
\begin{figure}[t]
\includegraphics[angle=270,scale=0.5]{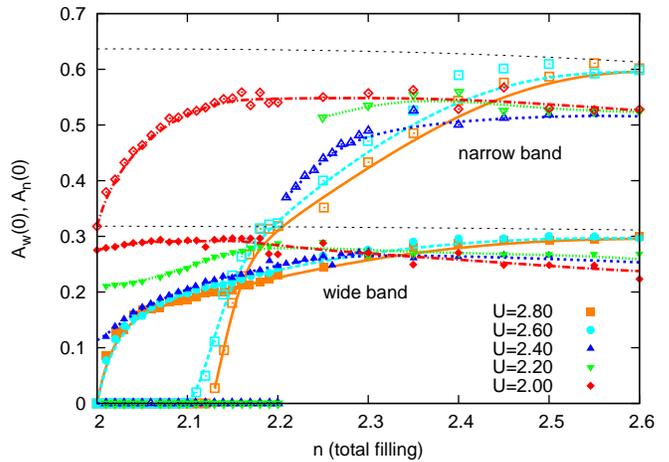}
\caption{(Color online) Spectral weight at the Fermi level for the two-band model. The thin dashed curves represent the spectral weights at the Fermi level according to Luttinger's theorem. All other curves are guides to the eyes only.
}\label{fig:N0}
\end{figure}

From our MEM-results for the spectral function we determine in particular the spectral weight at the Fermi level. Our results for this quantity, both for the narrow and for the wide band, are presented in Fig.\ \ref{fig:N0}. 
Within DMFT, M\"uller-Hartmann\cite{ZPB.76.211} showed for {\it single-band\/} models that, in the ground state, the spectral weight at the Fermi level is not renormalized by the interaction as a consequence of Luttinger's theorem. His proof is also applicable to multiband Hamiltonians, diagonal in the band- and spin-indices, such as the $J_z$-model (\ref{eq:model}) considered here. The reason is that, for such Hamiltonians, the multiband Green function, the Weiss field and the self-energy are also diagonal in the band- and spin-indices. Hence, Luttinger's theorem applies also to the multiband case, if the self-energy has quasiparticle properties: $\mathrm{Im}\,\Sigma(\omega) = {\cal O}(\omega^2)\ (\omega\rightarrow 0)$ on the real axis. Similar arguments can be found in Ref.\ \onlinecite{held_et_al_phys_stat_sol_review}. For comparison, we have also plotted, in Fig.\ \ref{fig:N0}, the spectral weight at the Fermi level as predicted by Luttinger's theorem (dashed lines).

It is clear from Fig.\ \ref{fig:N0} that the agreement between the spectral weights $A_\mathrm{w}(0)$ for the wide band and $A_\mathrm{n}(0)$ for the narrow band, both calculated at the Fermi level, and the values predicted by Luttinger's theorem improves with doping. In particular, there are significant deviations at low doping between the data and the reference curves.  There are several reasons for this: First of all, the narrow band is non-metallic and, hence, not a Fermi liquid at low doping (and for $U\gtrsim 2.1$).
Secondly, the data were sampled at finite temperature and the reference curve is valid only at $T=0$. Third, one expects that the bands, even when they are metallic, retain some non-Fermi-liquid properties at interaction values which, at half filling, are characteristic of an OSMP. Quantitatively, one deduces from Fig.\ \ref{fig:N0} that the metallicity (and hence the quasiparticle properties) of the wide and the narrow band develop in the doping ranges $2.0 \lesssim n \lesssim 2.3$ and $2.3 \lesssim n \lesssim 2.6$, respectively. For larger doping concentrations and interactions $U\gtrsim 2.6$, the agreement between the data and the reference curves is quite good.


\subsection{\label{subsec:QP}Self-energy and quasiparticle weights}
\begin{figure}
\includegraphics[angle=270,scale=0.5]{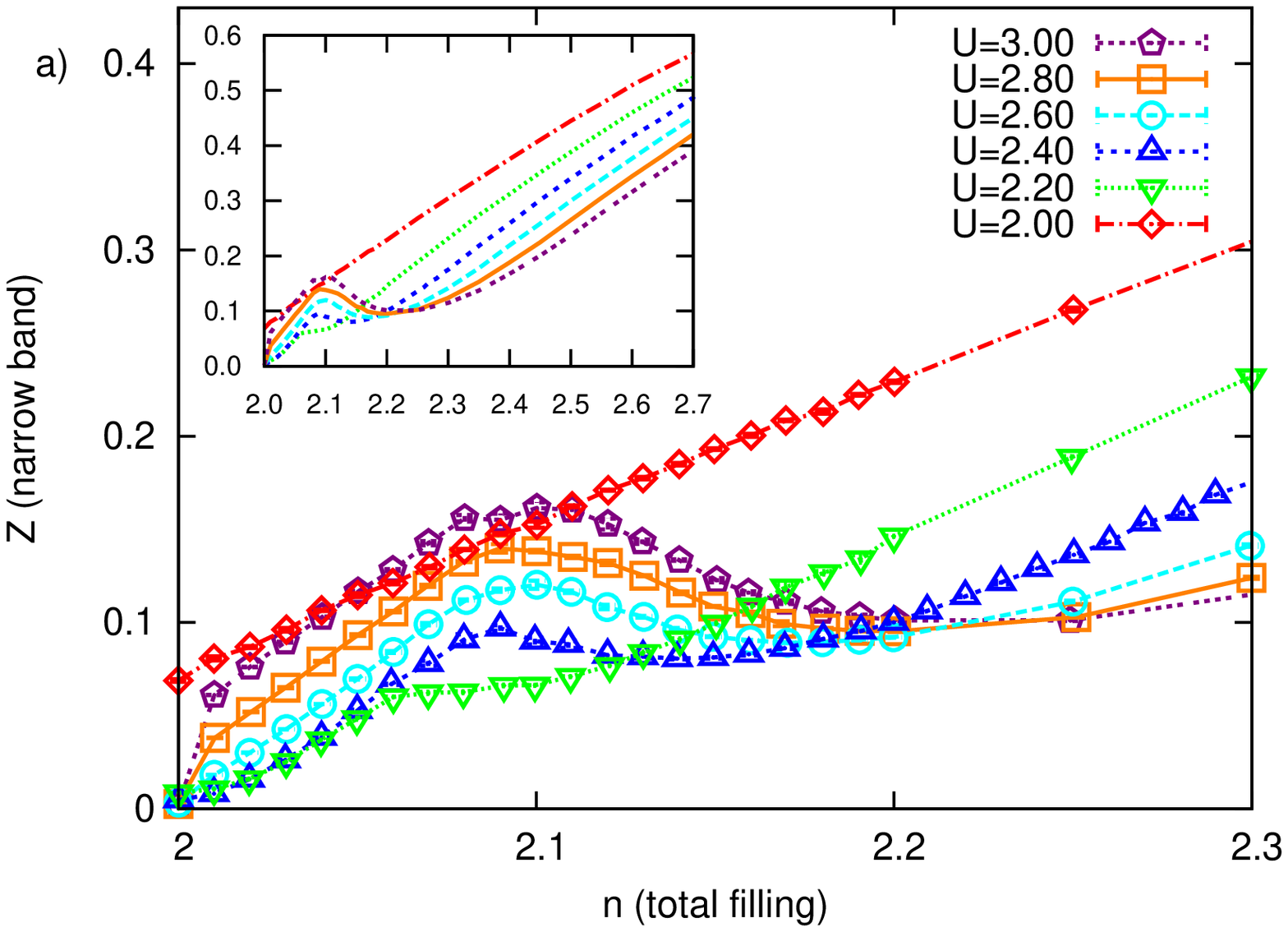}\\
\includegraphics[angle=270,scale=0.5]{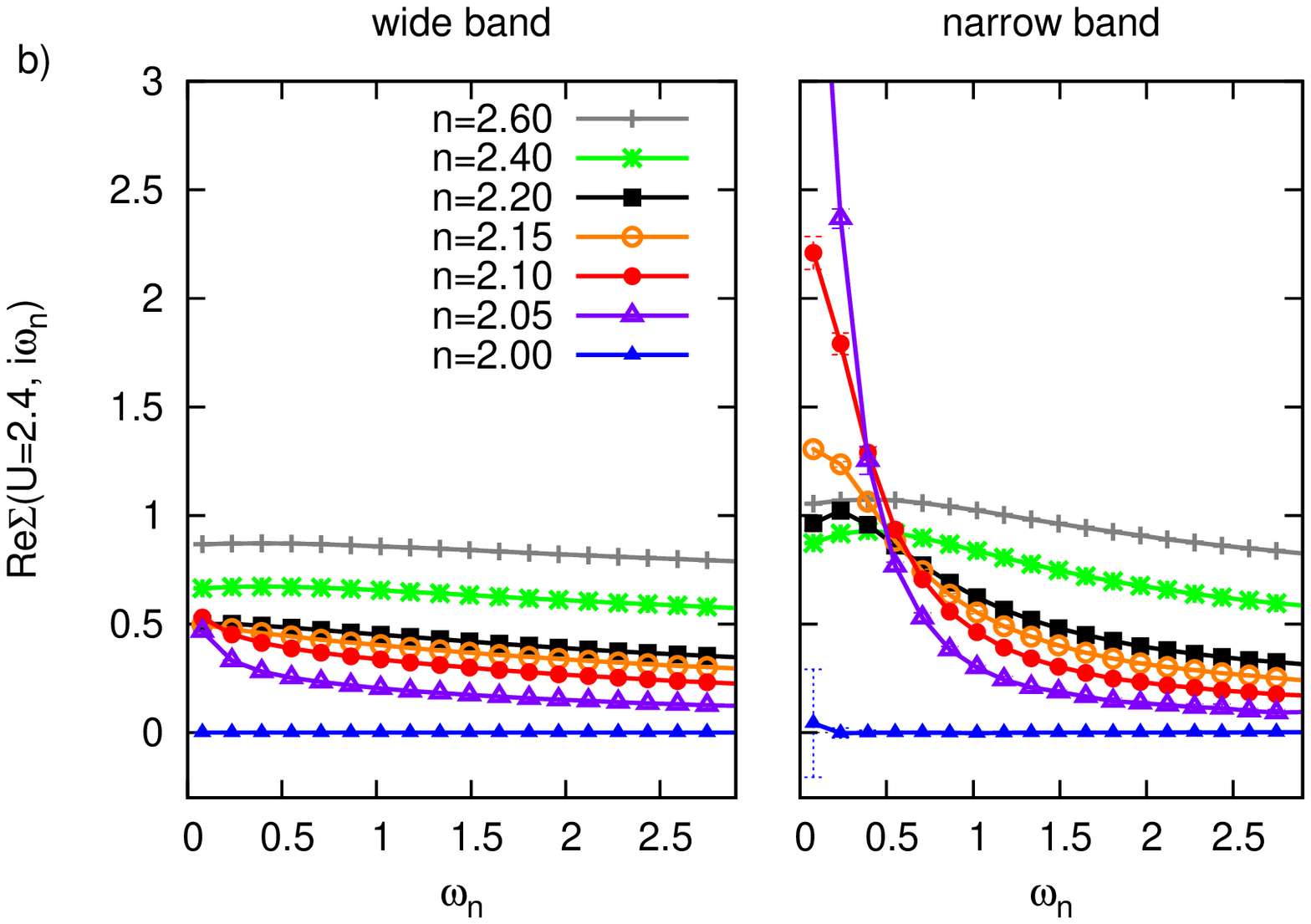}
\caption{(Color online) ($a$) Quasiparticle weight $Z_{\text{n}}$ as defined in Eq. (\ref{eq:Z}) for the narrow band in the two-band model. $Z_{\text{n}}$ develops a local maximum as a function of filling  in the weakly doped regime $2.05\lesssim n\lesssim 2.15$ with increasing interaction $U$. There are no further anomalies at stronger doping (see inset). ($b$) Real part of the self-energy calculated for $U=2.4$ at discrete Matsubara frequencies for the two-band model. The self-energy exhibits an anomaly (i.e., a strong enhancement at small frequencies) in the intermediate doping range ($2.05\lesssim n\lesssim 2.2$). The lines are guides to the eyes only.
}\label{fig:QP}
\end{figure}

We briefly discuss the behavior of the quasiparticle weight (QPW) $Z$, as defined by Eq.\ (\ref{eq:Z}), as a function of interaction and density. As stressed before, discrete estimate (\ref{eq:Z}) of $Z$ can, strictly speaking, only be interpreted as a physical ``quasiparticle weight'' if both bands are Fermi liquids. If one now studies the behavior of the QPW in the two-band Hubbard-model away from half filling for interaction values {\it outside\/} the Fermi-liquid regime ($U\gtrsim 2$), one expects and indeed finds an anomaly: The QPW of the narrow band does {\it not\/} depend monotonically on the filling [Fig.\ \ref{fig:QP}a], as it would in a Fermi liquid phase, but is increasingly enhanced at $n\approx 2.1$ with increasing $U$. To make sure that the behavior in Fig.\ \ref{fig:QP}a is not a numerical artifact, these results were counter-checked with a different QMC solver.

Having established that the QPW behaves anomalously outside the Fermi liquid regime, it is interesting to search for possible non-Fermi-liquid anomalies in the full frequency-dependent self-energies, from which the QPWs have been computed according to Eq.\ (\ref{eq:Z}). The self-energies behave smoothly as a function of the (discrete) Matsubara frequencies, down to the first Matsubara frequency, for all metallic solutions. However, already for the moderately large interaction value $U=2.4$, a clear anomaly can be seen in the real part of the self-energy [Fig.\ \ref{fig:QP}b]. In the doping region between $n=2.00$ and $n=2.40$ some intermediate solutions show divergent behavior for $\omega_{n}\to 0$, which strongly differs from the smooth solutions for densities $n=2.00$ and $n=2.40$. Moreover, $\Sigma(i\omega_n)$ shows a non-monotonic dependence on the filling at the first four Matsubara frequencies $\omega_1$, $\omega_2$, $\omega_3$, and $\omega_4$. This contrasts strongly with the characteristics found in the metallic regime, where $Z(i\omega_n)$ is monotonic as a function of $n$. Hence the self-energies, too, show anomalous behavior, as was found before for the secant estimate [see Eq. (\ref{eq:Z})] or Fig.\ \ref{fig:QP}a.
Note that the real part of the self-energy has to vanish on the imaginary axis due to particle-hole symmetry at half filling, even in the insulating phase.

\begin{figure}
\includegraphics[angle=270,scale=0.5]{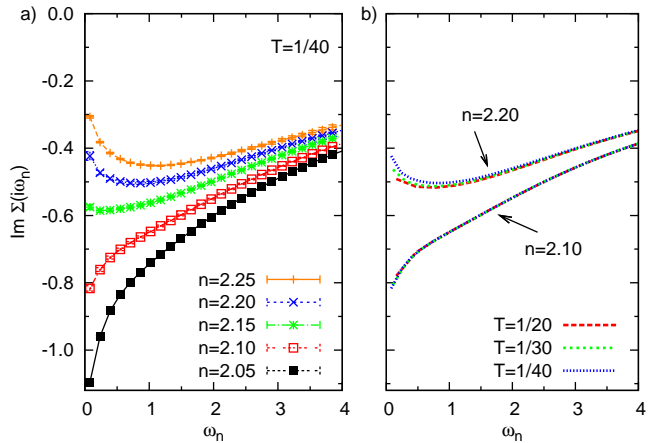}
\caption{(Color online) ($a$) Imaginary part of the self-energy $\mathrm{Im}\,\Sigma(\im \omega_n)$ for various total fillings $n$ at $U=2.4$ and $T=1/40$; intermediate minima are seen at low frequencies only for $n\ge 2.15$. ($b$) Results for $\mathrm{Im}\,\Sigma$ computed at different temperatures for selected densities (at $U=2.4$). The curves are hardly distinguishable for $n=2.1$; for $n=2.2$, the low-frequency minimum becomes more pronounced with decreasing $T$.
}\label{fig:ImS}
\end{figure}

Let us now come back to the question of how closely non-Fermi-liquid properties are connected with orbital-selective Mott physics (for anisotropic Hund rule couplings). In Fig.\ \ref{fig:spectraW}, we have already seen that the low-frequency dip in the wide-band spectrum, characteristic of a non-FL, persists at least up to $n=2.1$, the boundary of the OSM phase; however, some traces remain up to about $n=2.15$. So it is not clear, at this point, whether the OSMP and the non-FL phase are really identical. In order to avoid the ambiguities of analytic continuation, we will now discuss this issue on the basis of the imaginary-time self-energy data shown in Fig.\ \ref{fig:ImS}. 
Subpanel (a) shows $\mathrm{Im}\,\Sigma(\im \omega_n)$ for $U=2.4$ and $T=1/40$ (i.e., the same parameters as in Fig.\ \ref{fig:spectraW}) across a range $2.05\le n\le 2.25$ of densities. Close to half filling, at $n\le 2.1$, $\mathrm{Im}\,\Sigma(\im \omega)$ is a monotonic function of $\omega$ which clearly tends toward finite non zero values (of order 1) for $\omega\to 0$, i.e., shows characteristic non-FL behavior. In contrast, at $n=2.25$, far away from half filling (and from the OSMP), the self-energy clearly decays at small frequencies, consistent with the Fermi liquid scenario $\mathrm{Im}\,\Sigma(\im \omega) \propto {\cal O}(\omega) + {\cal O}(T^2)$. In between, the situation is less clear: it is impossible to decide, from this data alone, whether the apparent finite value to which the data extrapolates for $\omega\to 0$ represents thermal or intrinsic non-FL effects. Therefore, we have checked the impact of variations of the temperature for two selected densities. As seen in Fig.\ \ref{fig:ImS} (b), the curves connecting the discrete Matsubara data are on top of each other for $n=2.1$. Evidently, the low-temperature physics is determined by genuine non-FL scattering in this case; temperature effects are negligible. In contrast, a clear trend with decreasing temperature toward a Fermi-liquid form is seen for $n=2.2$; however, much lower temperatures would be needed in order to verify the full restoration of FL properties which would be very costly using QMC. This is even more true for $n=2.15$ for which $\mathrm{Im}\,\Sigma(\im \omega_n)$ shows only a very shallow minimum. So we have established that Fermi liquid properties are, indeed, restored for large enough doping; unfortunately our data cannot show whether this boundary matches the narrow-band OSMT at low temperatures. However, we have clearly shown that the wide band is a non-FL throughout the orbital-selective Mott phase. Note that the present Matsubara-frequency analysis nicely confirms the conclusions drawn from the wide-band spectra [Fig. 8] and, thereby, the reliability of the maximum-entropy procedure.


\section{\label{summary}Summary and Conclusion}

We first summarize our results. In this paper we studied electronic correlations within two-band $J_z$ model (\ref{eq:model}) using a high-precision QMC-DMFT code, both close to and far away from half filling.

We calculated both static and dynamic properties. Among the static properties, we focused particularly on band-specific fillings and intraorbital double occupancies as functions of interaction and doping. The most important result for these quantities is their pinning at the Mott insulating state for the narrow band. Our results for static quantities in a wide range of interaction and doping values were summarized in a phase diagram. In the section on the dynamical properties we focused on the frequency dependence of the spectral function and, in particular, on the density of states at the Fermi level. We calculated the density of states, both of the wide and of the narrow band, and compared the results to those for a single-band system. The main finding here is that, in the two-band model, the density of states is somewhat broadened, and that (compared to single-band spectra) weight from the Kondo resonance is shifted into the dips between the resonance and the Hubbard bands. A physical explanation for this is that the interorbital coupling introduces additional scattering channels for the electrons, which reduce quasiparticle lifetimes and, hence, smear out the spectrum.

In this paper, we followed the common practice in the study of
orbital-selective Mott transitions and focused on \emph{paramagnetic} phases,
excluding magnetic states. Still, the true thermodynamic equilibrium states
of model (\ref{eq:model}) may well be magnetic for the parameters of
interest. In fact, it is known for the half-filled case (in high dimensions)
that the N\'{e}el temperature at $U_{c1}$ is about \emph{six} times larger than
the critical temperature $T_{c1}\approx0.02$ of the OSMT (\onlinecite{pvd2007})
so that the OSMTs observed in paramagnetic calculations should be hidden
inside the antiferromagnetic equilibrium state. However, studies of 
symmetry-broken states are notoriously difficult away from half-filling
due to the large variety of possibilities. Also, various mechanisms
exist which can suppress ordered phases, such as disorder, spin-orbit coupling or
longer-range hopping.\cite{pvd2007} In particular, \emph{finite
dimensionality} can be expected to suppress magnetic states effectively
in planar systems such as $\mathrm{Ca_{2-x}Sr_xRuO_4}$. Therefore, and since
Mott transitions have been observed\cite{Koga05PRB,Gorelik09} to persist as narrow crossovers far above 
critical temperatures, results obtained in the paramagnetic phase should be qualitatively valid,
at least at intermediate temperatures.

To conclude, we demonstrated that, for interaction values in the
OSMP range $2\lesssim U\lesssim 2.6$, the entire doping range
$0\leq n\leq 4$ is subdivided into three different regimes by the
behavior of the narrow band. At low doping, both the particle
number and the double occupancy of the narrow band are pinned to
their Mott insulating values, so that, upon increasing the doping,
the wide band absorbs all additional electrons or holes; in this regime, 
the wide band shows characteristic non-Fermi-liquid behavior. In the
intermediate doping regime, i.e., from a critical doping
concentration $n_{\mathrm{crit}}(U)$ onward, the narrow band
leaves the pinned state, so that both bands contribute to the
compressibility and become metallic. Third, for a nearly filled
(or nearly empty) system with $n\gtrsim 3.6$ (or $n\lesssim 0.4$),
the narrow band is effectively a band insulator (or effectively
empty), so that, also in this regime, only the wide band
contributes to the compressibility. In this third regime,
correlation effects become surpressed and quantum fluctuations
vanish. Surprisingly, in the double occupancy this effect can
qualitatively already be seen for densities, not too far away from
half filling ($|n-2|\gtrsim 0.5$). Taken together, our findings
give a fairly complete and reliable description of the physics of two-band model (\ref{eq:model}) with inequivalent bands, in
which genuine multi-band and orbital-selective features are
clearly separated from generic correlation effects (that also
exist in single-band models).

\vspace{1ex}
{\bf Acknowledgments --}
The work of one of us (E.J.) was supported by the Graduate School of Excellence ``Materials Science in Mainz'', funded by the German Research Foundation with both federal and state support within the framework of the Excellence Initiative.


\bibliography{two_band.bib}

\end{document}